\documentclass[twocolumn,amsmath,amssymb,floatfix,aps,prb,shownopacs,footinbib,superscriptaddress]{revtex4-1}
\usepackage{amsmath,amssymb,bm,graphicx,url,epsf,color}
\usepackage{natbib} 
\usepackage[british]{babel}
\usepackage{wasysym}
\usepackage{MnSymbol}
\setcitestyle{super} 

\begin{document}

\title{Electronic heat flow and thermal shot noise in quantum circuits}

\author{E.~Sivre}
\email{These authors contributed equally to this work.}
\affiliation{Universit\'e Paris-Saclay, CNRS, Centre de Nanosciences et de Nanotechnologies (C2N), 91120 Palaiseau, France}
\author{H.~Duprez}
\email{These authors contributed equally to this work.}
\affiliation{Universit\'e Paris-Saclay, CNRS, Centre de Nanosciences et de Nanotechnologies (C2N), 91120 Palaiseau, France}
\author{A.~Anthore}
\affiliation{Universit\'e Paris-Saclay, CNRS, Centre de Nanosciences et de Nanotechnologies (C2N), 91120 Palaiseau, France}
\affiliation{Universit\'{e} de Paris, C2N, 91120 Palaiseau, France}
\author{A.~Aassime}
\affiliation{Universit\'e Paris-Saclay, CNRS, Centre de Nanosciences et de Nanotechnologies (C2N), 91120 Palaiseau, France}
\author{F.D.~Parmentier}
\affiliation{Universit\'e Paris-Saclay, CNRS, Centre de Nanosciences et de Nanotechnologies (C2N), 91120 Palaiseau, France}
\author{A.~Cavanna}
\affiliation{Universit\'e Paris-Saclay, CNRS, Centre de Nanosciences et de Nanotechnologies (C2N), 91120 Palaiseau, France}
\author{A.~Ouerghi}
\affiliation{Universit\'e Paris-Saclay, CNRS, Centre de Nanosciences et de Nanotechnologies (C2N), 91120 Palaiseau, France}
\author{U.~Gennser}
\affiliation{Universit\'e Paris-Saclay, CNRS, Centre de Nanosciences et de Nanotechnologies (C2N), 91120 Palaiseau, France}
\author{F.~Pierre\thanks{frederic.pierre@c2n.upsaclay.fr}}
\email[e-mail: ]{frederic.pierre@c2n.upsaclay.fr}
\affiliation{Universit\'e Paris-Saclay, CNRS, Centre de Nanosciences et de Nanotechnologies (C2N), 91120 Palaiseau, France}

\maketitle

{\sffamily 
When assembling individual quantum components into a mesoscopic circuit, the interplay between Coulomb interaction and charge granularity breaks down the classical laws of electrical impedance composition.
Here we explore experimentally the thermal consequences, and observe an additional quantum mechanism of electronic heat transport.
The investigated, broadly tunable test-bed circuit is composed of a micron-scale metallic node connected to one electronic channel and a resistance.
Heating up the node with Joule dissipation, we separately determine, from complementary noise measurements, both its temperature and the thermal shot noise induced by the temperature difference across the channel (`delta-$T$ noise').
The thermal shot noise predictions are thereby directly validated, and the electronic heat flow is revealed.
The latter exhibits a contribution from the channel involving the electrons' partitioning together with the Coulomb interaction.
Expanding heat current predictions to include the thermal shot noise, we find a quantitative agreement with experiments.
}

\begin{figure}[!htb]
\renewcommand{\figurename}{\textbf{Figure}}
\renewcommand{\thefigure}{\textbf{\arabic{figure}}}
\centering\includegraphics[width=\columnwidth]{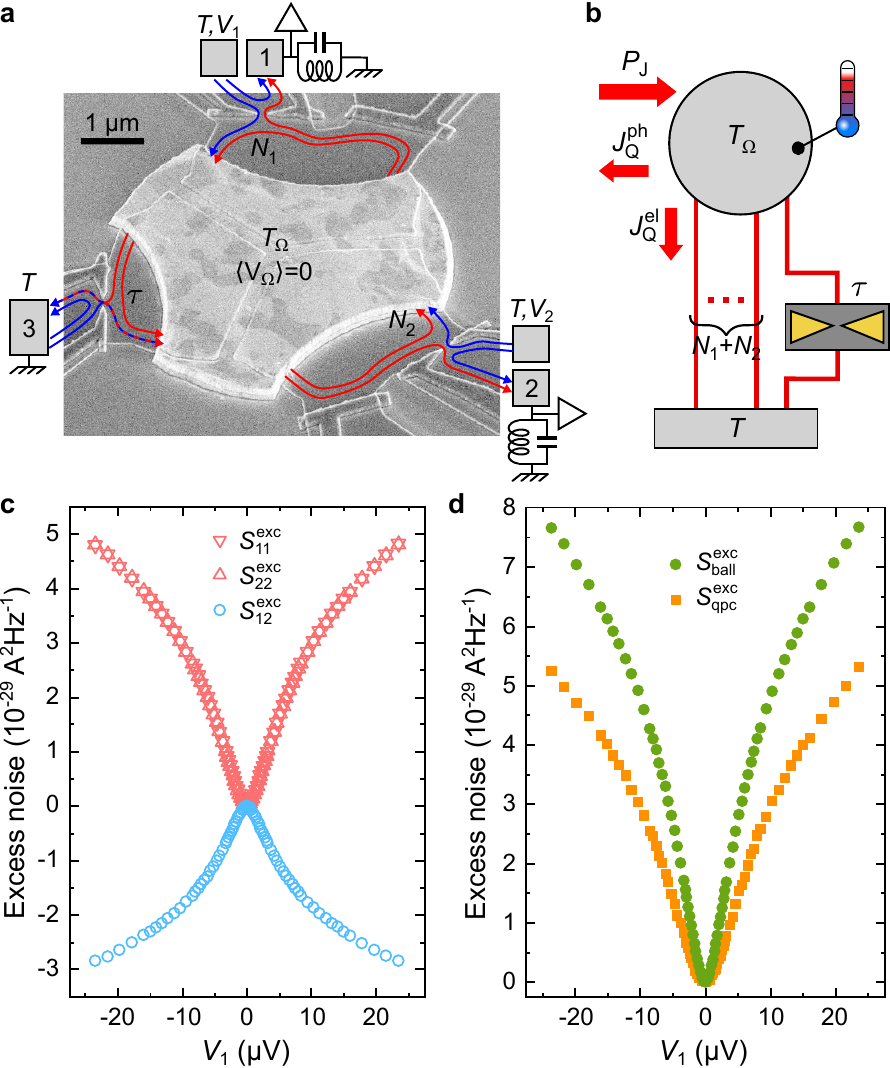}
\caption{
\footnotesize
Experimental approach.
\textbf{a}, Device e-beam micrograph with measurement setup schematic.
A single generic channel of arbitrary electron transmission probability $\tau$, as well as $N_1$ and $N_2$ ballistic (perfectly transmitted) channels, are separately connected to a small metallic island.
\textbf{b}, Schematic heat balance representation between injected Joule power ($P_\mathrm{J}$) and outgoing heat currents, from electrons to phonons ($J_\mathrm{Q}^\mathrm{ph}$) and through the connected electronic channels ($J_\mathrm{Q}^\mathrm{el}$).
\textbf{c}, Excess auto- and cross-correlation measurements versus $V_1=-V_2$, in the illustrative configuration $N=2$ ($N_1=N_2=1$), $\tau\sim0.5$.
\textbf{d}, Extracted excess noise sources per ballistic channel ($S_\mathrm{ball}^\mathrm{exc}$) and across the generic single-channel quantum point contact of transmission $\tau$ ($S_\mathrm{qpc}^\mathrm{exc}$), from the data in (c).
\normalsize
}
\label{fig1}
\end{figure}

Heating generally drives the crossover from quantum to classical behaviors; nevertheless, heat itself is ruled by quantum mechanics.
In recent years, experimental explorations of quantum thermal phenomena have been emerging at a rising pace\cite{Schwab2000,Giazotto2012,Ronzani2018}.
In particular, the quantum of thermal conductance, a universal basic building block of heat quantum transport, is now firmly established for bosons\cite{Schwab2000,Timofeev2009}, fermions\cite{Jezouin2013b,Srivastav2019} and quasiparticles that may be anyons\cite{Banerjee2017}, as well as up to macroscopic\cite{Partanen2016} and room temperature\cite{Mosso2017,Cui2017} scales.
However, despite the strong influence of Coulomb interaction on electricity in small quantum circuits\cite{SCT1992,Likharev1999,Parmentier2011,Jezouin2016}, its impact on the quantum transport of heat remains barely explored experimentally\cite{Dutta2017,Sivre2018,Banerjee2018}.
In a first step for perfectly ballistic circuits, where there is no back-scattering along any of the connected electronic channels, a recent observation\cite{Sivre2018} was made of the predicted\cite{Slobodeniuk2013} heat Coulomb blockade taking place without any concomitant reduction of the electrical conductance.
In this limit and at low temperatures, the Coulomb interaction manifests itself as the systematic suppression of a single channel for the evacuation of heat from a small circuit node\cite{Slobodeniuk2013,Sivre2018}.
Here we address elementary quantum circuits including one generic electronic channel, of arbitrary electron transmission probability.
An unexpected increase in the flow of heat is observed and quantitatively accounted for by an additional quantum heat transport mechanism, involving the association of thermal shot noise (dubbed `delta-$T$ noise' in Ref.~\citenum{Lumbroso2018}) and Coulomb interaction.

We obtain the heat current\textendash temperature characteristics by controllably injecting a dc power into a small floating circuit node connecting a quantum channel to a linear resistance, and by monitoring in-situ the resulting increase in the electrons' temperature.
A complication is that the partition of electrons in the generic channel breaks the Johnson-Nyquist proportionality between excess noise and node temperature increase\cite{Blanter2000,Lumbroso2018}, which was previously used for the thermometry of ballistic circuits\cite{Jezouin2013b,Banerjee2017,Banerjee2018,Sivre2018,Srivastav2019}.
We overcome this difficulty with an experimental procedure involving complementary measurements of both the auto- and cross-correlations of electrical fluctuations.
This provides us, separately, with the local electronic temperature in the metallic node, as well as with the thermal shot noise.
The latter is found in good agreement with predictions derived within the scattering approach\cite{Martin1992,Blanter2000}, in which Coulomb effects have been encapsulated in the temperature-dependent conductance (reduced by the dynamical Coulomb blockade\cite{SCT1992}).
The node temperature increase, both in terms of injected power and electron transmission probability across the channel, exposes an additional heat current contribution involving thermal shot noise.

\vspace{\baselineskip}
{\large\noindent\textbf{Results}}\\
{\noindent\textbf{Test-bed for electronic channels in dissipative environments.}}
An e-beam micrograph of the device is shown in Fig.~1a together with a schematic representation of the measurement setup.
The small floating circuit node that is heated is materialized by the central micron-scale metallic island (in brighter grey), of separately characterized self-capacitance $C\simeq3.1$\,fF.
It is in essentially perfect electrical contact with a standard Ga(Al)As two-dimensional electron gas underneath the surface.
The 2D gas is immersed in a perpendicular magnetic field corresponding to the integer quantum Hall regime at filling factor two.
In this regime, the current flows along two adjacent quantum Hall edge channels depicted by lines with arrows indicating the propagation direction. 
Three quantum point contacts (QPC) are formed in the 2D electron gas by applying negative voltages on surface split gates coupled capacitively.
A single (spin-polarized) short electronic channel of tunable transmission probability $\tau\in [0,1]$ is implemented at the left QPC.
The top and right QPCs are tuned to a different, ballistic regime:
they are set to fully transmit, respectively, $N_1$ and $N_2$ channels forming together an adjustable linear resistance\cite{Jezouin2013,Anthore2018} $R=R_\mathrm{K}/N$, with $R_\mathrm{K}=h/e^2$ the electrical resistance quantum ($h$ the Planck constant, $e$ the electron charge) and $N=N_1+N_2$.
Further away, the quantum Hall channels are connected to large electrodes at base temperature $T\simeq8$\,mK, represented in Fig.~1a by grey rectangles.

\vspace{\baselineskip}
{\noindent\textbf{Electronic heat flow determination.}}
The electrons within the central island are heated to $T_\Omega$ by dissipating a known Joule power $P_\mathrm{J}\simeq(N_1V_1^2+N_2V_2^2)/2R_\mathrm{K}$, with $V_1$ ($V_2$) the voltage applied to the top (right) large electrode (Methods).
The island's dc voltage is pinned to $\langle V_\Omega \rangle=0$, by imposing $N_1V_1=-N_2V_2$, such that the generic channel experiences a pure temperature bias $T_\Omega-T$ without dc voltage.
As illustrated in Fig.~1b, energy conservation in the stationary regime implies $P_\mathrm{J}=J_\mathrm{Q}^\mathrm{el}+J_\mathrm{Q}^\mathrm{ph}$, with $J_\mathrm{Q}^\mathrm{el}$ the heat flow across the connected electronic channels and $J_\mathrm{Q}^\mathrm{ph}$ the heat transferred from the electrons within the island to the phonons.
In practice, electron-phonon heat transfers are negligible only for $T_\Omega\lesssim20$\,mK\cite{Sivre2018}.
However, as $J_\mathrm{Q}^\mathrm{ph}$ only depends on temperatures ($T_\Omega$, $T$), and not on the connected electronic channels ($\tau$, $N$), it can be calibrated by tuning the circuit to the ballistic regime ($\tau\in\{0,1\}$).
Using the previously established heat Coulomb blockade predictions for ballistic channels\cite{Slobodeniuk2013,Sivre2018}, we find that all the data with $\tau\in\{0,1\}$, $N\in\{2,3,4\}$ and $T\in\{8,16\}$\,mK can be accurately reproduced using the same $J_\mathrm{Q}^\mathrm{ph}\simeq2.7\times10^{-8}\left(T_\Omega^{5.7}-T^{5.7}\right)$\,W (Methods).
At intermediate transmission probability ($0<\tau<1$), the unknown electronic heat flow is then obtained by subtracting the above $J_\mathrm{Q}^\mathrm{ph}$ from the injected Joule power ($J_\mathrm{Q}^\mathrm{el}=P_\mathrm{J}-J_\mathrm{Q}^\mathrm{ph}$).

\vspace{\baselineskip}
{\noindent\textbf{Local temperature increase measurement.}}
The island's electronic temperature $T_\Omega$ is determined from the low-frequency (MHz) current fluctuations measured on the top (1) and right (2) large electrodes (Methods). 
The excess auto- and cross-correlation spectral density, from which the zero-bias offset is removed, are plotted in Fig.~1c versus $V_1$ for the illustrative configuration $N_1=N_2=1$ at $\tau\sim0.5$.
In a nutshell, combining these data gives us access separately to the current noise sources originating from the QPC hosting a single generic channel ($S_\mathrm{qpc}$) and from the ballistic channels ($S_\mathrm{ball}$ per channel), both shown in Fig.~1d.
This is possible because these two noise sources contribute with the same sign to the experimental auto-correlation signal, while with an opposite sign to the cross-correlation (Methods).
The temperature $T_\Omega$ is then obtained using solely the ballistic noise source $S_\mathrm{ball}$, directly resulting from the thermal fluctuations of the electronic states' population in the baths.
This robust connection manifests itself as a straightforward, and previously used\cite{Jezouin2013b,Banerjee2017,Banerjee2018,Sivre2018,Srivastav2019}, generalization of the fluctuation-dissipation relation for the thermal noise $S_\mathrm{ball}=4k_\mathrm{B}\bar{T}/R_\mathrm{K}$, where $\bar{T}=(T_\Omega+T)/2$ is the average temperature\cite{Blanter2000,Blanter2000chaotic}.
In practice, the excess noise data (with respect to $V_{1,2}=0$) gives us access to the temperature increase $T_\Omega-T$, while $T$ is separately measured (Methods).

\begin{figure*}[!tb]
\renewcommand{\figurename}{\textbf{Figure}}
\renewcommand{\thefigure}{\textbf{\arabic{figure}}}
\centering\includegraphics[width=180mm]{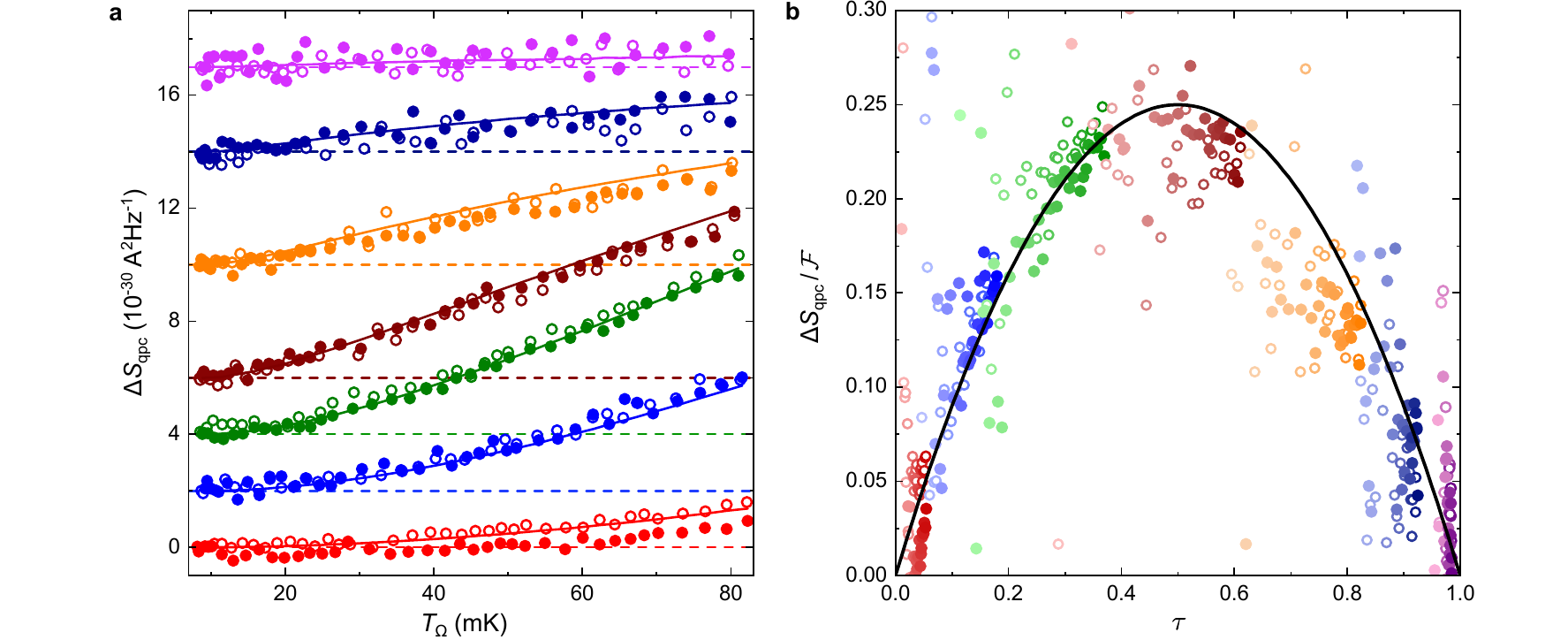}
\caption{
\footnotesize
Thermal shot noise (delta-$T$ noise).
\textbf{a}, Symbols represent the experimental QPC noise at $N=2$ from which the average thermal noise was removed ($\Delta S_\mathrm{qpc}=S_\mathrm{qpc}-4k_\mathrm{B}\bar{T}\tau/R_\mathrm{K}$, with $\bar{T}=(T_\Omega+T)/2$).
Measurements at different gate voltage tunings of the QPC are shifted vertically, with the applied offsets shown as horizontal dashed lines. 
Open and full symbols distinguish separate sequences of measurements.
Continuous lines display Eq.~\ref{eqSqpc} predictions.
\textbf{b}, The $\tau(1-\tau)$ partition signature is shown as a continuous line versus $\tau$.
Symbols represent $\Delta S_\mathrm{qpc}/\mathcal{F}$, where the $\tau$-independent function $\mathcal{F}(T_\Omega,T)$ is the predicted thermal shot noise's temperature dependence (see text).
A lighter (darker) symbol coloring indicates a low (large) $T_\Omega-T$ corresponding to a higher (lower) experimental uncertainty.
\normalsize
}
\label{fig2}
\end{figure*}

\vspace{\baselineskip}
{\noindent\textbf{Shot noise induced by a temperature difference.}}
Generic channels driven out-of-equilibrium are generally expected to exhibit, in addition to the average thermal noise, a shot noise induced by the electron partitioning into a transmitted electron and a reflected electron\cite{Martin1992,Blanter2000}.
In particular, the current noise spectral density at low frequencies ($\omega\ll k_\mathrm{B}T/\hbar$), for a single channel of transmission probability $\tau$, reads\cite{Blanter2000}:
\begin{equation}
S_\mathrm{qpc}^\mathrm{thy}=\frac{4k_\mathrm{B}\bar{T}\tau}{R_\mathrm{K}}+\frac{2\tau(1-\tau)}{R_\mathrm{K}}\int dE\,\left[f_\mathrm{T_\Omega}(E)-f_\mathrm{T}(E)\right]^2,\label{eqSqpc}
\end{equation}
with $f_\mathrm{T,T_\Omega}(E)$ the Fermi distributions in the connected baths at different temperatures and/or voltages.
The average thermal noise and the shot noise are, respectively, the first and second term on the right-hand side of Eq.~\ref{eqSqpc}.
Whereas the shot noise induced by either a voltage difference or a frequency irradiation is experimentally well-established (see references in Ref.~\citenum{Blanter2000} and also Ref.~\citenum{Reydellet2003}), the thermal shot noise resulting from the partition of electrons in the sole presence of a temperature difference was observed only recently in Ref.~\citenum{Lumbroso2018}, where it is referred to as `delta-$T$ noise'.
Although convincing, this observation did not allow for a one-to-one comparison of the individual data points with the theory, because the possibly multiple electronic channels were incompletely characterized by the measurement of their parallel conductance.
In contrast, in the present work with a single generic channel, the QPC conductance $G_\mathrm{qpc}=\tau e^2/h$ completely determines the transmission probability $\tau$.
In Fig.~2a, following Ref.~\citenum{Lumbroso2018}, we focus on the thermal shot noise (delta-$T$ noise) $\Delta S_\mathrm{qpc}$ obtained by removing the average Johnson-Nyquist noise ($\Delta S_\mathrm{qpc}=S_\mathrm{qpc}-4k_\mathrm{B}\bar{T}\tau/R_\mathrm{K}$).
The $\Delta S_\mathrm{qpc}$ data at $N_1=N_2=1$ (symbols) are plotted versus $T_\Omega$ for several gate voltage tunings of the single channel QPC.
The predictions (continuous lines), calculated without any adjustable parameter using Eq.~\ref{eqSqpc}, closely match the data (for control experiments, see Supplementary Fig.~1 at other $\{N_1,N_2\}$ and Supplementary Fig.~2 at a larger base temperature $T\simeq16\,$mK).
Note that the simultaneously measured $G_\mathrm{qpc}=\tau e^2/h$ depends on the temperatures $T$ and $T_\Omega$, because of the quantum back-action of the series $RC$ circuit\cite{Parmentier2011} also referred to as the dynamical Coulomb blockade\cite{SCT1992}.
Remarkably, we find that the effect of Coulomb interaction is accurately encapsulated, at experimental resolution, into the renormalized $\tau$ injected in Eq.~\ref{eqSqpc}.
Figure~2b directly reveals the partition origin of the shot noise induced by a temperature difference.
The data points represent this experimental shot noise renormalized by the predicted, $\tau$-independent temperature function $\mathcal{F}(T_\Omega,T)=(2/R_\mathrm{K})\int dE\,\left[f_\mathrm{T_\Omega}(E)-f_\mathrm{T}(E)\right]^2$.
The good agreement observed between $\Delta S_\mathrm{qpc}/\mathcal{F}$ and $\tau(1-\tau)$ attests of the underlying partition mechanism.

\begin{figure}[!htb]
\renewcommand{\figurename}{\textbf{Figure}}
\renewcommand{\thefigure}{\textbf{\arabic{figure}}}
\centering\includegraphics[width=\columnwidth]{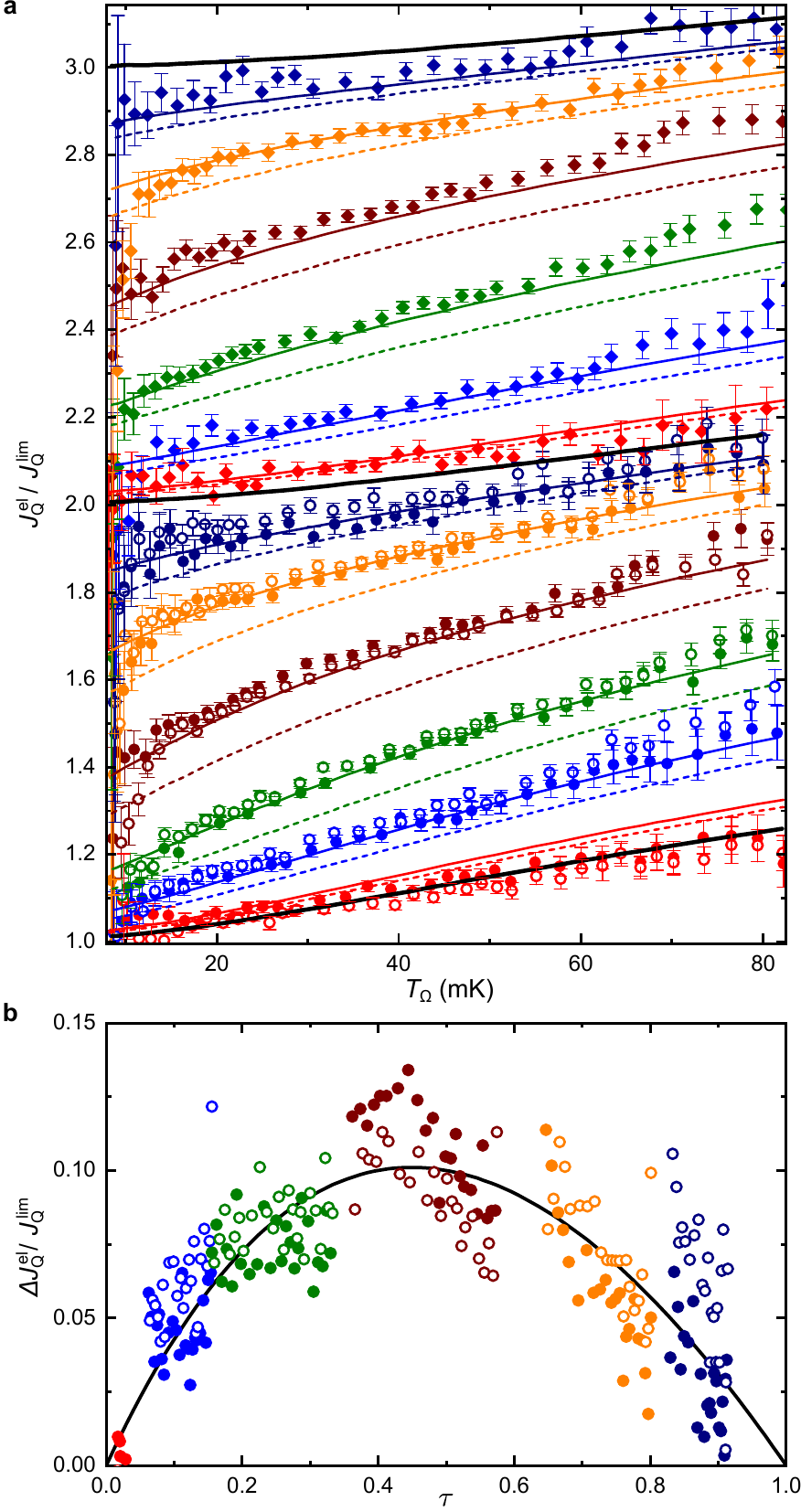}
\caption{
\footnotesize
Electronic heat flow.
\textbf{a}, Experimental $J_\mathrm{Q}^\mathrm{el}/J_\mathrm{Q}^\mathrm{lim}$ (with $J_\mathrm{Q}^\mathrm{lim}=\pi^2k_\mathrm{B}^2(T_\Omega^2-T^2)/6h$) are plotted as symbols versus $T_\Omega$ with $N=2$ (circles) and $N=3$ (diamonds), for a broad range of QPC tunings (colors).
Error bars represent the standard statistical error.
Black continuous lines are predictions at $\tau=0$ for $N=2$ (bottom), 3 (middle) and 4 (top).
Dashed lines are interpolations between ballistic predictions, linear in the measured $\tau$.
Continuous lines are theoretical predictions.
\textbf{b}, Symbols represent the difference $\Delta J_\mathrm{Q}^\mathrm{el}$ between experimental $J_\mathrm{Q}^\mathrm{el}$ ($N=2$, $T_\Omega\in[17,65]$\,mK in panel (a)) and the corresponding interpolation between ballistic predictions, normalized by $J_\mathrm{Q}^\mathrm{lim}$.
The continuous line displays versus $\tau$ the low-temperature prediction $\tau(1-\tau)/(N+\tau)$ for $N=2$.
\normalsize
}
\label{fig3}
\end{figure}

\vspace{\baselineskip}
{\noindent\textbf{Electronic heat flow from a small quantum circuit node.}}
We now address the electronic flow of heat across the QPC and ballistic channels.
In conductors, the thermal conductance $G_\mathrm{Q}$ is frequently found to be directly proportional to the electrical conductance $G_\mathrm{el}$, through the so-called Wiedemann-Franz (WF) law $G_\mathrm{Q}=\mathcal{L}G_\mathrm{el}$ with $\mathcal{L}=\pi^2k_\mathrm{B}^2/3e^2$ the Lorenz number.
While this relation holds between the quantum of thermal and electrical conductances, it generally breaks down in quantum circuits assembled from several interconnected channels.
In particular, it was shown that the thermal conductance from a small, heated circuit node connected by ballistic channels is reduced from the WF expectation by precisely one quantum of thermal conductance at low temperatures\cite{Slobodeniuk2013,Sivre2018}, whatever the total number of channels.
With such a fixed reduction, the increment by $\mathcal{L}/R_\mathrm{K}$ of the thermal conductance when adding an extra ballistic channel (starting from at least one) nevertheless follows the WF relation.
Is this also the case if the electrical conductance is increased continuously, by sweeping the transmission probability across an electronic channel from $\tau=0$ to 1?
The answer is no, as we will now show.

Figure~3a exhibits as symbols, versus $T_\Omega$, the experimental electronic heat flow $J_\mathrm{Q}^\mathrm{el}$ normalized by the quantum limit per channel $J_\mathrm{Q}^\mathrm{lim}=\pi^2k_\mathrm{B}^2(T_\Omega^2-T^2)/6h$, for different circuit settings spanning the full range of $\tau$ at both $N=2$ and $N=3$ (see Supplementary Fig.~3b for $N=4$, and Supplementary Fig.~4 for a control experiment at $T\simeq16$\,mK).
The three thick black continuous lines display the full, temperature dependent heat Coulomb blockade prediction for 2 (bottom), 3 (middle) and 4 (top) ballistic channels\cite{Slobodeniuk2013} (Methods).
Note the small, predicted deviations developing with temperature from the complete heat Coulomb blockade of a single channel ($J_\mathrm{Q}^\mathrm{el}/J_\mathrm{Q}^\mathrm{lim}=N-1$) that fully applies only in the limit of low temperatures $T_\Omega,T\ll\hbar/k_\mathrm{B}RC$.
Open and full circles (full diamonds) are data points obtained for $N=2$ ($N=3$) ballistic channels, with different settings of the generic channel encoded by different colors.
The dashed lines represent linear interpolations between ballistic predictions at $N$ and $N+1$ weighted, respectively, by $1-\tau$ and $\tau$ measured for the compared data (same color).
For example, the brown dashed line in the top part of Fig.~3a (closest to $J_\mathrm{Q}^\mathrm{el}/J_\mathrm{Q}^\mathrm{lim}\sim2.5$) is given by $\tau(T_\Omega)$ times the prediction for three ballistic channels (thick black line near $J_\mathrm{Q}^\mathrm{el}/J_\mathrm{Q}^\mathrm{lim}\sim2$, Methods) plus $1-\tau(T_\Omega)$ times the prediction for four ballistic channels (thick black line near $J_\mathrm{Q}^\mathrm{el}/J_\mathrm{Q}^\mathrm{lim}\sim3$), with $\tau(T_\Omega)$ the renormalized conductance simultaneously measured during the acquisition of the top brown data points of corresponding $T_\Omega$ (in practice a linear interpolation is performed between discrete measurements of $\tau(T_\Omega)$).
The difference between dashed lines and data points is particularly significant at intermediate $\tau$.
This shows that the thermal conductance increase does not reduce to a linear, WF-like function of the electrical conductance.
In contrast, quantitative predictions based on the Langevin approach of Ref.~\citenum{Slobodeniuk2013} but including the partition noise from the generic channel (colored continuous lines, Methods) lie close to the data, without any adjustable parameter.
At low temperatures $T_\Omega,T\ll\hbar/k_\mathrm{B}RC$, the difference between theory (thy) predictions $J_\mathrm{Q}^\mathrm{thy}$ and the WF extension (linear in $\tau$) of heat Coulomb blockade predictions for ballistic channels $(N+\tau-1)\times J_\mathrm{Q}^\mathrm{lim}$, reads:
\begin{equation}
J_\mathrm{Q}^\mathrm{thy}-(N+\tau-1)\times J_\mathrm{Q}^\mathrm{lim}\simeq \frac{\tau(1-\tau)}{N+\tau}\times J_\mathrm{Q}^\mathrm{lim}.\label{eqJQthymWFlowT}
\end{equation}
Note that $J_\mathrm{Q}^\mathrm{thy}=0$ for $N=0$ at low temperatures, whatever the value of $\tau$ (see Refs.~\citenum{Duprez2019b,Idrisov2018,Clerk2001} for the electrons' state preservation concomitant to the absence of heat transfers).
The $\tau(1-\tau)$ numerator attests of the role of electron partition in this additional heat transport mechanism.
We also point out that this heat current contribution vanishes at higher temperatures, when Coulomb effects become negligible (Methods).
This shows straightforwardly the essential role of Coulomb interaction, which combines with electron partition into a different form of quantum heat transport.
Figure~3b provides direct experimental evidences for an underlying partition mechanism (see also Supplementary Fig.~3a,c), by subtracting from the renormalized electronic heat flow at $N=2$ (symbols in Fig.~3a) the corresponding WF (linear) interpolation (dashed lines in Fig.~3a).
Focusing here on the temperature range $T_\Omega\in[17,65]$\,mK where measurements are most accurate (see error bars in Fig.~3a), a convincing agreement is found with $\tau(1-\tau)/(2+\tau)$ plotted as a continuous line versus $\tau$.

\vspace{\baselineskip}
{\large\noindent\textbf{Discussion}}\\
We have experimentally investigated the heat flow and thermally induced shot noise, or delta-$T$ noise\cite{Lumbroso2018}, in an elementary quantum circuit composed of one small metallic node (island) connected by several ballistic channels and by one generic electronic channel of arbitrary electron transmission probability.
Applying a temperature bias, without dc voltage across the generic channel, we measured the thermal shot noise and determined the overall electronic heat flow from the island. 
The former is found in direct quantitative agreement with thermal shot noise predictions in the scattering approach\cite{Blanter2000}, computed using the simultaneously measured transmission probability (note that a different prediction has recently been made for the thermal shot noise in the fractional quantum Hall regime\cite{Rech2020}).
The latter displays an additional heat flow contribution.
The underlying mechanism involves in particular the Coulomb charging energy of the island, which effectively freezes its total charge at low temperatures and thereby induces correlations between the heat carrying electrical current fluctuations propagating along the connected channels\cite{Slobodeniuk2013} (Methods).
In a fully ballistic circuit (without thermal shot noise), these correlations amount to the recently observed systematic blockade of a single channel for the flow of heat, independently of the total number of channels\cite{Slobodeniuk2013,Sivre2018}.
Here, with a generic channel, a thermal shot noise is impinging on the island and fractionalized among all the outgoing channels by the frozen island charge imposed by Coulomb interaction\cite{Idrisov2019}.
This combination of Coulomb interaction and thermal shot noise underpins the presently observed additional heat transport mechanism (Methods).

Advancing our understanding of the mechanisms of quantum heat transport and establishing the thermal shot noise contribution is essential for exploiting heat and noise to unveil exotic physics\cite{Altimiras2012,Inoue2014a,Banerjee2018}, and is bound to play a role in the thermal and signal to noise management of future quantum devices.
The present work also demonstrates measurement strategies widening the range of experimental systems eligible for thermal explorations:
by exploiting complementary auto- and cross-correlation measurements of the electrical fluctuations, we have shown that the different sources of noise can be accessed separately.
We expect that such advanced combinations of fluctuation measurements will play an increasing role in the thermal and noise investigations of quantum circuits.

\vspace{\baselineskip}
{\large\noindent\textbf{Methods}}\\
\footnotesize
{\noindent\textbf{Sample.}} 
The Al(Ga)As 2DEG has an electron density of $2.5.10^{11}\,$cm$^{-2}$, a mobility of $10^6\,$cm$^2$V$^{-1}$s$^{-1}$ and is located 105\,nm below the surface. 
The central island is formed from a metallic layering of nickel (30\,nm), gold (120\,nm) and germanium (60\,nm), which is thermally annealed at 440$^\circ$C for 50\,s to make an electrical contact with the 2DEG.
The two quantum Hall edge channels at filling factor $\nu=2$ are found in near perfect contact with the island, with a reflection probability below $6\,10^{-3}$ (see Methods in Ref.~\citenum{Jezouin2016} for a detailed description of the characterization procedure).
The short $\sim1\,\mu$m distance between metallic island and QPC combined with the low temperatures ($T_\Omega\lesssim80\,$mK) ascertains that the interaction between co-propagating channels can be safely ignored (see e.g. Ref.~\citenum{leSueur2010}), as in previous works with the same sample\cite{Iftikhar2015,Iftikhar2016,Jezouin2016,Anthore2018,Iftikhar2018}.
The self-capacitance of the island $C\simeq3.1$\,fF (corresponding to a charging energy $E_\mathrm{C}=e^2/2C\approx k_\mathrm{B}\times0.3\,$K) is obtained from standard Coulomb diamond measurements (with all channels connected to the device tuned in the tunnel regime).\\

{\noindent\textbf{Noise measurement setup.}} 
The time-dependent current fluctuations $\delta I_1(t)$ and $\delta I_2(t)$ impinging, respectively, on electrodes 1 and 2 are first amplified with a cryogenic amplifier located on the 4\,K stage of a dilution refrigerator, and with a room temperature amplifier.
They are then digitized at 10\,Mbit/s and sent to a computer.
The Fourier auto- and cross-correlations analysis are performed over a 180\,kHz bandwidth centered on $0.855$\,MHz (the resonant frequency of the $LC$ oscillators shown in Fig.~1a).
The amplification gains $G_{1,2}^\mathrm{amp}$ are separately calibrated from the same standard shot-noise vs voltage bias measurements used to determine the base temperature $T$ (see corresponding section). 
We find that $G_{1,2}^\mathrm{amp}$ are stable along each run, but slightly different from cooldown to cooldown. 
Averaging 862 (2840) shot noise vs voltage bias sweeps, the statistical uncertainty on $G_{1,2}^\mathrm{amp}$ is below $0.09\%$ ($0.04\%$) for the first (second) experimental run shown here.
The cross-correlation gain $G_\mathrm{X}^\mathrm{amp}$ is also impacted by the matching between the two resonators.
For a perfect match, $G_\mathrm{X}^\mathrm{amp}=\sqrt{G_{1}^\mathrm{amp}G_{2}^\mathrm{amp}}$.
In general, a correction factor $c_{12}$ needs to be introduced $G_\mathrm{X}^\mathrm{amp}=\sqrt{G_{1}^\mathrm{amp}G_{2}^\mathrm{amp}}\times c_{12}$.
This factor $c_{12}$ is experimentally characterized at $\tau=0$ ($N_{1,2}\neq0$) from the robust relation $\Delta S_{11}=\Delta S_{22}=-\Delta S_{12}$, which directly results from the negligible charge accumulation on the island at the measurement frequencies.
In practice, we find an essentially perfect resonators' match ($c_{12}\approx 1.000$ and $0.993$ for the first and second cooldown, respectively).\\

{\noindent\textbf{Dissipated Joule power.}} 
The bulk of the Joule power dissipated within the electronic fluid in the metallic island is given by the expression $P_\mathrm{J}\simeq (N_1V_1^2+N_2V_2^2)/2R_\mathrm{K}$.
We also include the small additional contributions $P_\mathrm{J}^\mathrm{ac}$ due to the extra power dissipated from the small ac voltages $V_{1,2,3}^\mathrm{ac}\simeq 0.23\,\mu\mathrm{V_{rms}}$ applied (at different low frequencies) to the three source electrodes (to simultaneously measure with lock-in the conductances across each of the three QPCs), as well as a separately characterized small triboelectric voltage from the pulse tube vibrations specifically developing on the source electrode 1 (feeding the top QPC) $V_1^\mathrm{tribo}\simeq 0.4\,\mu\mathrm{V_{rms}}$: 
\begin{align}
P_\mathrm{J}^\mathrm{ac}=&\frac{1}{2R_\mathrm{K}(N+\tau)}\times \big[ \{(V_{1}^\mathrm{ac})^2+(V_1^\mathrm{tribo})^2\}N_1(N_2+\tau)\nonumber \\
&+(V_{2}^\mathrm{ac})^2N_2(N_1+\tau) +(V_{3}^\mathrm{ac})^2\tau N\big]. \label{eqPinjac}
\end{align}
In practice, $P_\mathrm{J}^\mathrm{ac}\in[2,6]$\,aW is below 1\% of $P_\mathrm{J}$ at $T_\Omega\gtrsim20\,$mK.
It corresponds to a temperature increase in the island of $\sim0.3\,$mK at zero dc bias (see next section -- Base electron temperature).
Note that we avoid possible mismatch from the thermoelectric voltage developing along the measurement lines by applying a current dc bias.
It is converted onchip into a voltage exploiting the well-defined quantum Hall resistance $R_\mathrm{K}/\nu$ connecting current biased electrodes and cold electrical grounds.\\

{\noindent\textbf{Base electron temperature.}} 
The base electronic temperature $T$ is extracted from standard shot-noise measurements, applying a dc bias voltage directly to a QPC set to a transmission probability of one half, with the floating island bypassed using side gates (see Methods in Ref.~\citenum{Iftikhar2016} for further details).

Due to the small $P_\mathrm{J}^\mathrm{ac}$ (see above section -- Dissipated Joule power), the temperature of the floating island is slightly higher than $T$ even in the absence of a dc voltage.
This small temperature increase is obtained by measuring the cross-correlations at zero dc bias $V_1=V_2=0$ (carefully calibrating instrumental offsets just before and after each measurement sequence), from the relation:
\begin{equation}
T_\Omega(V_{1,2}=0)-T\simeq-\frac{R_\mathrm{K}}{2k_\mathrm{B}}\frac{N+\tau}{N_1N_2}S_{12}(V_{1,2}=0),
\end{equation}
which straightforwardly relies on the generalized fluctuation-dissipation relation.
Although there are deviations from the generalized fluctuation-dissipation relation in the presence of a generic channel, as studied in this work, this approximation is excellent for small $T_\Omega(V_{1,2}=0)-T\ll T$ such as in the present case.
We find $T_\Omega(V_{1,2}=0)-T\sim0.3$\,mK (always below $0.6$\,mK), consistent with expectations based on the value of $P_\mathrm{J}^\mathrm{ac}$ given by Eq.~\ref{eqPinjac}.
This small temperature difference is included in the experimental determination of $T_\Omega$.\\

\begin{figure}[!htb]
\renewcommand{\figurename}{\textbf{Figure}}
\renewcommand{\thefigure}{\textbf{\arabic{figure}}}
\centering\includegraphics[width=\columnwidth]{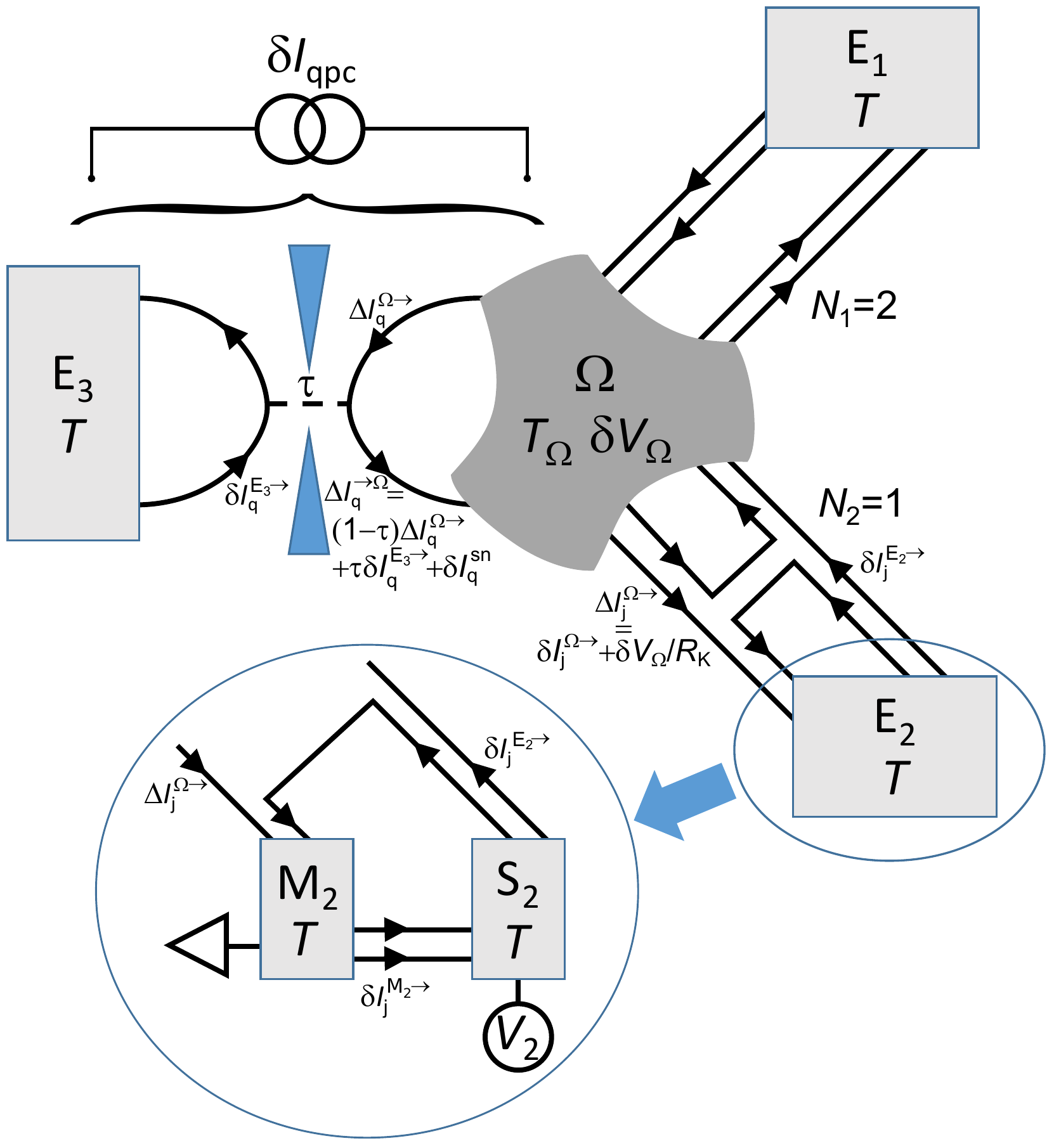}
\footnotesize
\caption{
Noise schematic.
Graphical representation of the different current and voltage fluctuations discussed in the text.
}
\label{figED_SSchem}
\end{figure}

{\noindent\textbf{Excess electron temperature and shot noise.}} 
This section details how are obtained the excess electron temperature, $\Delta T_\Omega=T_\Omega-T_\Omega(V_{1,2}=0)$, and the resulting excess noise generated across the generic QPC, $S_\mathrm{qpc}^\mathrm{exc}=\langle\delta I_\mathrm{qpc}^2\rangle-\langle\delta I_\mathrm{qpc}^2\rangle(V_{1,2}=0)$.
A schematic representation of the circuit is shown in Fig.~\ref{figED_SSchem} with arrows indicating the chirality also corresponding to the convention used for positive currents.
The large electrodes labeled $E_\mathrm{n}$ ($n\in\{1,2\}$) include each a measurement electrode $M_\mathrm{n}$ and a voltage biased source electrode $S_\mathrm{n}$.
The floating central metallic node is labeled $\Omega$.

First, let us separately consider a current fluctuation $\delta I_\mathrm{qpc}$ generated across the generic QPC (see Fig.~\ref{figED_SSchem}), and determine the resulting current fluctuations $\delta I_\mathrm{M1,M2}^\mathrm{qpc}$ impinging on the measurement electrodes $M_{1,2}$.
As the corresponding charge accumulated in the island relaxes very fast compared to the measurement frequencies ($1/R_\mathrm{K}C\sim10\,\mathrm{GHz}\gg1$\,MHz), the current $\delta I_\mathrm{qpc}$ injected in the island is compensated by the outgoing current from the resulting voltage fluctuation $\delta V_\Omega$ of the floating island.
This reads $\delta I_\mathrm{qpc}=(N+\tau) \delta V_\Omega/R_\mathrm{K}$ (for a treatment of charge relaxation at arbitrary frequencies see section Heat Coulomb blockade predictions).
Then, $\delta I_\mathrm{M1(2)}^\mathrm{qpc}=N_{1(2)} \delta V_\Omega/R_\mathrm{K}=\delta I_\mathrm{qpc}N_{1(2)}/(N+\tau)$.
Consequently, the QPC noise's contribution to the auto-correlation signal is:
\begin{equation}
S_{11(22)}^\mathrm{qpc}=S_\mathrm{qpc}N_{1(2)}^2/(N+\tau)^2,
\end{equation}
and its contribution to the cross-correlation signal is:
\begin{equation}
S_{12}^\mathrm{qpc}=S_\mathrm{qpc}N_1N_2/(N+\tau)^2.
\end{equation}

Second, we separately consider a current fluctuation $\delta I_\mathrm{j}^{\Omega\rightarrow}$ emitted from the island, by the thermal fluctuations of electronic states' population at $T_\Omega$, into a ballistic channel $j$.
From the fast charge relaxation of the island pointed out in the previous paragraph, one obtains $\delta V_\Omega/R_\mathrm{K}=-\delta I_\mathrm{j}^{\Omega\rightarrow}/(N+\tau)$.
On the one hand, the current fluctuation measured on the electrode $M_1$ if the channel $j$ propagates toward the electrode $M_2$ ($j\in M_2$) is then $\delta I_\mathrm{M1}^{\Omega\rightarrow2}=-N_1/(N+\tau)\delta I_\mathrm{j\in M2}^{\Omega\rightarrow}$.
The corresponding autocorrelation signal on $M_1$, resulting from the thermal current fluctuations emitted toward $M_2$ (summing all $j\in M_2$) therefore reads:
\begin{equation}
S_{11}^{\Omega\rightarrow2}=\frac{N_1^2}{(N+\tau)^2}\sum_{j=1}^{N_2}\langle(\delta I_\mathrm{j}^{\Omega\rightarrow})^2\rangle=N_2\times\frac{N_1^2}{(N+\tau)^2} \langle(\delta I^{\Omega\rightarrow})^2\rangle,
\end{equation}
where the unimportant channel index $j$ is omitted in $\langle(\delta I^{\Omega\rightarrow})^2\rangle\equiv\langle(\delta I^{\Omega\rightarrow}_\mathrm{j})^2\rangle$ (independent of $j$).
On the other hand, the current fluctuation measured on the electrode $M_1$ if the channel $j$ is also connected to the electrode $M_1$ ($j\in M_1$) includes both the direct term $\delta I_\mathrm{j\in M1}^{\Omega\rightarrow}$ and the smaller $\delta V_\Omega$ contribution: 
$\delta I_\mathrm{M1}^{\Omega\rightarrow1}=\left[1-N_1/(N+\tau)\right]\delta I_\mathrm{j\in M1}^{\Omega\rightarrow}$. 
As a result
\begin{align}
S_{11}^{\Omega\rightarrow1}&=\left(1-\frac{N_1}{N+\tau}\right)^2\sum_{j=1}^{N_1}\langle(\delta I_\mathrm{j}^{\Omega\rightarrow})^2\rangle \nonumber\\
 &= N_1\times\left(1-\frac{N_1}{N+\tau}\right)^2\langle(\delta I^{\Omega\rightarrow})^2\rangle,
\end{align}
and
\begin{equation}
S_{12}^{\Omega\rightarrow1}=-N_1\times\left(1-\frac{N_1}{N+\tau}\right)\frac{N_2}{N+\tau} \langle(\delta I^{\Omega\rightarrow})^2\rangle.
\end{equation}

Summing up the independent contributions from the QPC ($\delta I_\mathrm{qpc}$) and from all ballistic channels (emitted $\delta I_\mathrm{j}^{\Omega\rightarrow}$ and absorbed $\delta I_\mathrm{j}^{E_n\rightarrow}$), one straightforwardly obtains for the auto-correlation signal:
\begin{align}
S_{11(22)}=&N_{1(2)}\left[\left(1-\frac{N_{1(2)}}{N+\tau}\right)^2+\frac{N_1N_2}{\left(N+\tau\right)^2}\right] \langle(\delta I^{\Omega\rightarrow})^2\rangle \nonumber\\
&+\frac{N_{1(2)}^2}{\left(N+\tau\right)^2}S_\mathrm{qpc} +\frac{N_{1(2)}^2N}{\left(N+\tau\right)^2} \langle(\delta I^{E\rightarrow})^2\rangle +S_\mathrm{offset1(2)},\label{eqSauto}
\end{align}
with $S_\mathrm{offset1(2)}$ a noise offset mostly corresponding to the amplification chain, and also including the thermal noise along the $2-N_{1(2)}$ reflected channels and along the $2$ quantum Hall channels propagating from measurement (M) to source (S) contacts (for the experimental bulk filling factor $\nu=2$; see Fig.~\ref{figED_SSchem}).
Similarly, one gets for the cross-correlation signal:
\begin{align}
S_{12}=&\frac{N_1N_2}{\left(N+\tau\right)^2} \big[ -(N+2\tau)\langle(\delta I^{\Omega\rightarrow})^2\rangle + S_\mathrm{qpc} + N\langle( \delta I^{E\rightarrow})^2\rangle \big].\label{eqScross}
\end{align}

Focusing on the excess signal with respect to $V_{1,2}=0$, one obtains from Eqs.~\ref{eqSauto} and \ref{eqScross}:
\begin{equation}
S_\mathrm{ball}^\mathrm{exc}=\frac{S_{11}^\mathrm{exc}}{2N_1}+\frac{S_{22}^\mathrm{exc}}{2N_2}-\frac{S_{12}^\mathrm{exc}N}{2N_1N_2},
\end{equation}
with $S_\mathrm{ball}^\mathrm{exc}=\langle(\delta I^{\Omega\rightarrow})^2\rangle-\langle(\delta I^{\Omega\rightarrow})^2\rangle(V_{1,2}=0)$ the excess noise generated across one ballistic channel.
From the Johnson-Nyquist-type relation well established in the ballistic case\cite{Slobodeniuk2013,Jezouin2013b,Banerjee2017,Banerjee2018,Sivre2018,Srivastav2019} $\langle(\delta I^{\Omega\rightarrow})^2\rangle=2k_\mathrm{B}T_\Omega/R_\mathrm{K}$, the excess island's temperature reads:
\begin{equation}
\Delta T_\Omega=\frac{R_\mathrm{K}}{2k_\mathrm{B}}\left(\frac{S_{11}^\mathrm{exc}}{2N_1}+\frac{S_{22}^\mathrm{exc}}{2N_2}-\frac{S_{12}^\mathrm{exc}N}{2N_1N_2}\right). \label{eqDTohm}
\end{equation}
Solving Eqs.~\ref{eqSauto} and \ref{eqScross} also provides $S_\mathrm{qpc}^\mathrm{exc}$:
\begin{equation}
S_\mathrm{qpc}^\mathrm{exc}=(N+2\tau)\left(\frac{S_{11}^\mathrm{exc}}{2N_1}+\frac{S_{22}^\mathrm{exc}}{2N_2}\right)+S_{12}^\mathrm{exc}\frac{(N+\tau)^2+\tau^2}{2N_1N_2}. \label{eqDSqpc}
\end{equation}
\\

{\noindent\textbf{Heat Coulomb blockade predictions.}} 
In this section we derive the predictions shown as continuous lines in Fig.~3 and Supplementary Figs.~3,4, for the electronic flow of heat $J_\mathrm{Q}^\mathrm{el}$ in the presence of a generic quantum channel.
We follow the Langevin approach developed for ballistic systems in Ref.~\citenum{Slobodeniuk2013}, and expand it to the case where the current is partially reflected with a probability $1-\tau$ on a quantum point contact inserted along one of the channels (the other channels remaining ballistic, see schematic in Fig.~\ref{figED_SSchem}).
The three main differences with Ref.~\citenum{Slobodeniuk2013} are: (i) the symmetry between channels is broken, (ii) a partition noise emerges at the generic quantum point contact, (iii) the transmission probability $\tau$ depends on the temperatures due to dynamical Coulomb blockade.

The heat flow $J_\mathrm{Qj}^\rightarrow$ propagating in one direction ($\rightarrow$) along one electronic channel ($j$) is obtained from the time dependent electrical current fluctuations $\Delta I_\mathrm{j}^\rightarrow$ propagating in the same direction at the considered location\cite{Slobodeniuk2013}:
\begin{equation}
J_\mathrm{Qj}^\rightarrow=\frac{\hbar}{2e^2}\int_{-\infty}^\infty\mathrm{d}\omega\left(\langle(\Delta I_\mathrm{j}^\rightarrow)^2\rangle-\langle(\Delta I_\mathrm{j}^\rightarrow)^2\rangle_\mathrm{vacuum}\right), \label{eqJQvsdI}
\end{equation}
with $\langle\rangle_\mathrm{vacuum}$ referring to the vacuum fluctuations at zero temperature.

If $\Delta I_\mathrm{j}^\rightarrow$ directly originates from the large, voltage biased electrodes ($S_{1,2,3}$ in $E_{1,2,3}$), then it only includes the emitted thermal current fluctuation $\delta I_\mathrm{j}^{E_n\rightarrow}$ (see Fig.~\ref{figED_SSchem}). 
These thermal fluctuations are assumed uncorrelated ($\langle \delta I_\mathrm{j}^{E_\mathrm{n}\rightarrow} \delta I_\mathrm{k}^{E_\mathrm{m}\rightarrow} \rangle=0$ for $j\neq k$ even at $m=n$) and of variance given by the usual thermal noise expression at the base temperature $T$\cite{Slobodeniuk2013}:
\begin{equation}
\langle (\delta I_\mathrm{j}^{E_\mathrm{n}\rightarrow})^2 \rangle (\omega) = \frac{\hbar\omega/R_\mathrm{K}}{-1+\exp\left[\hbar\omega/k_\mathrm{B}T\right]}.\label{eqdIT}
\end{equation}
Note the factor two difference with the standard low frequency expression $2k_\mathrm{B}T/R_\mathrm{K}$, in which the contribution at positive and negative frequencies are added.
Injecting Eq.~\ref{eqdIT} into Eq.~\ref{eqJQvsdI}, one obtains the usual expression $J_\mathrm{Qj}^\rightarrow=(\pi k_\mathrm{B}T)^2/6h$.

In contrast to the voltage biased electrodes, the floating metallic node's electrochemical potential exhibits fluctuations $\delta V_\Omega$ (related to charge fluctuations as e.g. in the voltage probe and dephasing probe models, see Ref.~\citenum{Blanter2000} and references therein).
These result in the emission of identical current fluctuations $\delta V_\Omega /R_\mathrm{K}$ in all outgoing channels\cite{Slobodeniuk2013,Blanter2000}.
Such current fluctuations add up with the thermal emission $\delta I_\mathrm{j}^{\Omega\rightarrow}$ of electrons from the central node: $\Delta I_\mathrm{j}^{\Omega\rightarrow}=\delta I_\mathrm{j}^{\Omega\rightarrow}+\delta V_\Omega/R_\mathrm{K}$, with $\langle \delta I_\mathrm{j}^{\Omega\rightarrow} \delta I_\mathrm{k}^{\Omega\rightarrow} \rangle=0$ for $j\neq k$ and a variance $\langle (\delta I_\mathrm{j}^{\Omega\rightarrow} )^2 \rangle$ given by the same Eq.~\ref{eqdIT} but with the island temperature $T_\Omega$ instead of $T$.
The integrand in Eq.~\ref{eqJQvsdI} therefore includes such correlations as $\langle \delta I_\mathrm{j}^{\Omega\rightarrow}\delta V_\Omega\rangle$.
These can be obtained from the connection to the island's charge fluctuations $\delta Q=C\delta V_\Omega$ ($\delta Q=Q-\langle Q\rangle$ with $Q$ the overall charge of the island, and $C$ its self-capacitance), which obey the charge conservation relation:
\begin{align}
i\omega\delta Q=&\sum_{j=1}^{N+1} (\Delta I_\mathrm{j}^{\rightarrow\Omega}-\Delta I_\mathrm{j}^{\Omega\rightarrow}) \nonumber\\
=& \Delta I_\mathrm{q}^{\rightarrow\Omega}-\delta I_\mathrm{q}^{\Omega\rightarrow}-\delta Q/R_\mathrm{K}C \nonumber\\
&+\sum_{j=1}^{N} (\delta I_\mathrm{j}^{E\rightarrow}-\delta I_\mathrm{j}^{\Omega\rightarrow})-N\delta Q/R_\mathrm{K}C,\label{eqQcons}
\end{align}
where we separated in the second equality the generic channel labeled with the index $q$ (first line) from the $N$ ballistic channels (second line).
In channel $q$, the incoming current fluctuations toward the island $\Delta I_\mathrm{q}^{\rightarrow\Omega}$ include three contributions:
\begin{equation}
\Delta I_\mathrm{q}^{\rightarrow\Omega}=\tau \delta I_\mathrm{q}^{E_{3}\rightarrow} + (1-\tau)\left( \delta I_\mathrm{q}^{\Omega\rightarrow} + \delta Q/R_\mathrm{K}C \right) + \delta I_\mathrm{q}^\mathrm{sn},\label{eqDIin}
\end{equation}
with the third term corresponding in the Langevin description to an uncorrelated noise source induced by the electrons' partition at the QPC.
At equilibrium ($T=T_\Omega$), the Johnson-Nyquist relation at low frequencies imposes $2\langle(\delta I_\mathrm{q}^\mathrm{sn})^2\rangle=\tau(1-\tau)\times4k_\mathrm{B}T/R_\mathrm{K}$ (the factor two is because positive and negative frequencies are included for this comparison).
In the non-equilibrium regime ($T\neq T_\Omega$), the information needed on $\delta I_\mathrm{q}^\mathrm{sn}$ for the heat current will be directly obtained from energy flow conservation at the input and output of the QPC (see below).
Note that we neglect in Eq.~\ref{eqDIin} the small time delay associated with the round loop path island-QPC-island (a delay of about $20$\,ps using a typical velocity of $10^5$\,m/s), and that the transmission probability $\tau$ is taken as a frequency independent value (that depends on $T$ and $T_\Omega$ due to dynamical Coulomb blockade, see e.g. Ref.~\citenum{Anthore2018}).
Combining Eqs.~\ref{eqQcons} and \ref{eqDIin} allows us to write $\delta Q$ as a function of uncorrelated noise sources:
\begin{align}
(i\omega+\tau_\Omega/R_\mathrm{K}C) \delta Q =&\tau (\delta I_\mathrm{q}^{E_3\rightarrow}-\delta I_\mathrm{q}^{\Omega\rightarrow})\nonumber\\
&+\delta I_\mathrm{q}^\mathrm{sn}+\sum_{j=1}^{N} (\delta I_\mathrm{j}^{E\rightarrow}-\delta I_\mathrm{j}^{\Omega\rightarrow}),\label{eqDQ}
\end{align}
where we introduced $\tau_\Omega$ defined as the sum of the transmission probabilities of the channels connected to the island:
\begin{equation}
\tau_\Omega=N+\tau.
\end{equation}
This straightforwardly makes it possible to formulate the integrands $\langle(\Delta I_\mathrm{q}^{\rightarrow\Omega})^2\rangle$ and $\langle(\Delta I_\mathrm{q}^{\Omega\rightarrow})^2\rangle$ as functions of uncorrelated noise sources (independently of $\delta V_\Omega$).
As an illustration, we obtain for the latter:
\begin{align}
\langle(\Delta I_\mathrm{q}^{\Omega\rightarrow})^2\rangle =& \frac{\langle(\delta  I_\mathrm{q}^\mathrm{sn})^2\rangle + \left(\tau_\Omega-\tau(1-\tau)\right)\langle(\delta I^{E\rightarrow})^2\rangle}{\tau_\Omega^2+(\omega R_\mathrm{K}C)^2} \nonumber\\
&+\left( 1 +\frac{\tau_\Omega-\tau(1-\tau)-2\tau\tau_\Omega}{\tau_\Omega^2+(\omega R_\mathrm{K}C)^2} \right) \langle(\delta I^{\Omega\rightarrow})^2\rangle,
\end{align}
where the arbitrary index $j$ is omitted.
The only missing ingredient is $\delta  I_\mathrm{q}^\mathrm{sn}$.
As pointed out above, the required information can be obtained most robustly from global heat conservation at the QPC: $J_\mathrm{Qq}^{E_3\rightarrow}+J_\mathrm{Qq}^{\Omega\rightarrow}=J_\mathrm{Qq}^{\rightarrow E_3}+J_\mathrm{Qq}^{\rightarrow\Omega}$, with $J_\mathrm{Qq}^{E_3\rightarrow}$ the flow of heat from the large electrode $E_3$ toward the QPC, $J_\mathrm{Qq}^{\Omega\rightarrow}$ the flow of heat from the island toward the QPC, $J_\mathrm{Qq}^{\rightarrow E_3}$ the flow of heat from the QPC toward $E_\mathrm{3}$, and $J_\mathrm{Qq}^{\rightarrow\Omega}$ the flow of heat from the QPC toward the island.
Using Eq.~\ref{eqJQvsdI}, this equality reads:
\begin{align}
\int_{-\infty}^\infty\mathrm{d}\omega\, &\langle(\delta  I_\mathrm{q}^\mathrm{sn})^2\rangle\times\left[ 1+\frac{\tau_\Omega-\tau(1-\tau)-2\tau\tau_\Omega}{\tau_\Omega^2+(\omega R_\mathrm{K}C)^2} \right] \nonumber\\
=& \int_{-\infty}^\infty\mathrm{d}\omega\, \tau(1-\tau)\left[ 1+\frac{\tau_\Omega-\tau(1-\tau)-2\tau\tau_\Omega}{\tau_\Omega^2+(\omega R_\mathrm{K}C)^2} \right] \nonumber \\
& \times \left\lbrace  \langle(\delta I_\mathrm{q}^{E_3\rightarrow})^2\rangle + \langle(\delta I_\mathrm{q}^{\Omega\rightarrow})^2\rangle \right\rbrace.
\end{align}\\
Summing up the contributions of all channels and performing the integration in Eq.~\ref{eqJQvsdI}, we obtain for the net heat flow from the metallic island:
\begin{align}
J_\mathrm{Q}^\mathrm{thy}=&\sum_{j=1}^{N+1} \left(J_\mathrm{Qj}^{\Omega\rightarrow}-J_\mathrm{Qj}^{\rightarrow\Omega}\right) \nonumber \\
=& \tau_\Omega\frac{\pi^2k_\mathrm{B}^2}{6h}(T_\Omega^2-T^2)-\tau_\Omega\frac{h(\tau_\Omega-\tau(1-\tau))}{(2\pi R_\mathrm{K}C)^2}\times\nonumber \\
&\left[ \Im \left(\frac{h\tau_\Omega/R_\mathrm{K}C}{2\pi k_\mathrm{B}T_\Omega }\right)- \Im \left(\frac{h\tau_\Omega/R_\mathrm{K}C}{2\pi k_\mathrm{B}T }\right) \right],\label{eqJQ}
\end{align}
with the function $\Im$ given by:
\begin{equation}
\Im(x)=\frac{1}{2}\left[ \ln\left(\frac{x}{2\pi}\right) - \frac{\pi}{x} -\psi\left( \frac{x}{2\pi} \right) \right],\label{eqI}
\end{equation}
with $\psi(z)$ the digamma function.
Equation~\ref{eqJQ} was used to calculate the predictions shown as continuous lines in Fig.~3a, Supplementary Fig.~3b and Supplementary Fig.~4.\\ 
At $\tau=0$ or 1, Eq.~\ref{eqJQ} reduces to the expression derived for a ballistic system\cite{Slobodeniuk2013} (see Methods in Ref.~\citenum{Sivre2018} for a similar formulation).
At high temperatures, Eq.~\ref{eqJQ} reduces to the non-interacting result matching the widespread Widemann-Franz law (without additional contribution from the partition noise):
\begin{align}
J_\mathrm{Q}^\mathrm{thy}\left(T,T_\Omega\gg\frac{\hbar\tau_\Omega}{k_\mathrm{B}R_\mathrm{K}C}\right)&\simeq \tau_\Omega\frac{\pi^2k_\mathrm{B}^2}{6h}(T_\Omega^2-T^2)\nonumber\\
&\simeq\tau_\Omega J_\mathrm{Q}^\mathrm{lim}.\label{eqJQhighT}
\end{align}
At low temperatures, Eq.~\ref{eqJQ} simplifies into:
\begin{align}
J_\mathrm{Q}^\mathrm{thy}\left(T,T_\Omega\ll\frac{\hbar\tau_\Omega}{k_\mathrm{B}R_\mathrm{K}C}\right)&\simeq \left(\tau_\Omega-1+\frac{\tau(1-\tau)}{\tau_\Omega}\right)\frac{\pi^2k_\mathrm{B}^2}{6h}(T_\Omega^2-T^2)\nonumber\\
&\simeq \left(\tau_\Omega-1+\frac{\tau(1-\tau)}{\tau_\Omega}\right)J_\mathrm{Q}^\mathrm{lim}.\label{eqJQlowT}
\end{align}
In this case, in addition to the systematic blockade of one ballistic channel ($-1$) with respect to the non-interacting case ($\tau_\Omega$), we find an additional contribution to the flow of heat whose partition character is signaled by the characteristic $\tau(1-\tau)$ dependence.\\

\begin{figure}[t]
\renewcommand{\figurename}{\textbf{Figure}}
\renewcommand{\thefigure}{\textbf{\arabic{figure}}}
\includegraphics[width=\columnwidth]{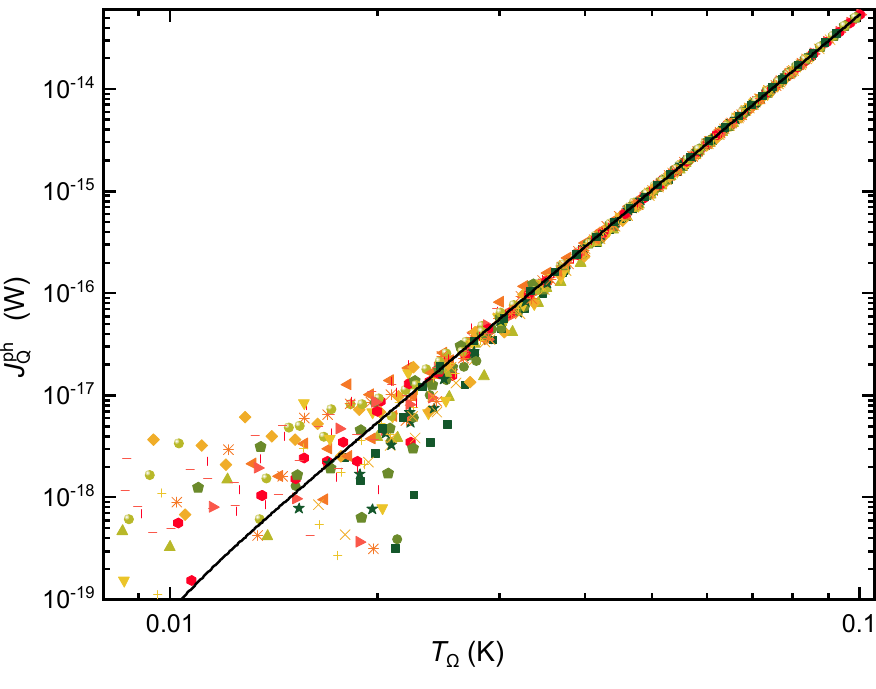}
\footnotesize
\caption{
Electron-phonon heat transfers.
Symbols represent the flow of heat $J_\text{Q}^\text{ph}$ from the electrons within the metallic island to the phonons at $T\simeq8\,$mK. 
Different symbols represent data points from different configurations $N\in \{2,3,4\}$ and $\tau\in\{0,1\}$, including measurements performed in two different cooldowns.
All the data collapse on $J_\text{Q}^\text{ph}=2.752\,10^{-8}(T_\Omega^{5.709}-T^{5.709})\,$W (continuous line).
}
\label{figED_fitphonons}
\end{figure}

{\noindent\textbf{Electron-phonon heat transfers.}} 
The figure~\ref{figED_fitphonons} displays the amount of heat transferred from electrons in the metallic island to cold phonons at base temperature $T\simeq8\,$mK.
It is obtained by subtracting from the injected Joule power $P_\mathrm{J}$ the known electronic heat flow $J_\mathrm{Q}^\mathrm{el}$ when the circuit is tuned in the ballistic regime (for the subtracted expression of $J_\mathrm{Q}^\mathrm{el}$, see Eq.~\ref{eqJQ} with $\tau\in\{0,1\}$ or Refs.~\citenum{Sivre2018,Slobodeniuk2013}).
The data from all ballistic configurations ($N\in \{2,3,4\}$, $\tau\in\{0,1\}$) collapse on the same curve, fitted by $J_\text{Q}^\text{ph}=\Sigma (T_\Omega^\alpha-T^\alpha)$ with $\Sigma=2.752\,10^{-8}\,\mathrm{W.K}^{-\alpha}$ and $\alpha=5.709$.
We checked that this power law also precisely accounts for $J_\text{Q}^\text{ph}$ at the larger temperature $T\simeq16\,$mK (data not shown).

\vspace{\baselineskip}
{\large\noindent\textbf{Data Availability}}\\
\footnotesize
The data that support the findings of this study are available from the corresponding author upon reasonable request.
\normalsize


\footnotesize
\vspace{\baselineskip}
{\normalsize\noindent\textbf{Acknowledgments}}\\
This work was supported by the French RENATECH network, the national French program Investissements d'Avenir (Labex NanoSaclay, ANR-10-LABX-0035) and the French National Research Agency (projects QuTherm, ANR-16-CE30-0010, and SIM-CIRCUIT, ANR-18-CE47-0014-01).

We thank Y.~Jin for providing the cryogenic HEMTs used for the noise measurements, and E.~Sukhorukov for fruitful discussions.

\vspace{\baselineskip}
{\normalsize\noindent\textbf{Author Contributions}}\\
E.S. and H.D. performed the experiment with inputs from A.Aa., A.An. and F.P.;
A.An., E.S., F.P. and H.D. analyzed the data and expanded the theory with help from A.Aa.;
F.D.P. fabricated the sample using developments from A.An.;
A.C., A.O. and U.G. grew the 2DEG;
F.P. led the project and wrote the manuscript with contributions from A.Aa., A.An., E.S., H.D. and U.G.

\vspace{\baselineskip}
{\normalsize\noindent\textbf{Competing Interests}}\\
The authors declare no competing interests. 

\vspace{\baselineskip}
{\normalsize\noindent\textbf{Materials \& Correspondence}}\\
Correspondence and requests for materials should be addressed to F.P. (frederic.pierre@c2n.upsaclay.fr).

\newpage
\setcounter{figure}{0}

\begin{figure*}[b]
\renewcommand{\figurename}{\textbf{Supplementary Figure}}
\renewcommand{\thefigure}{\textbf{\arabic{figure}}}
\includegraphics[width=\textwidth]{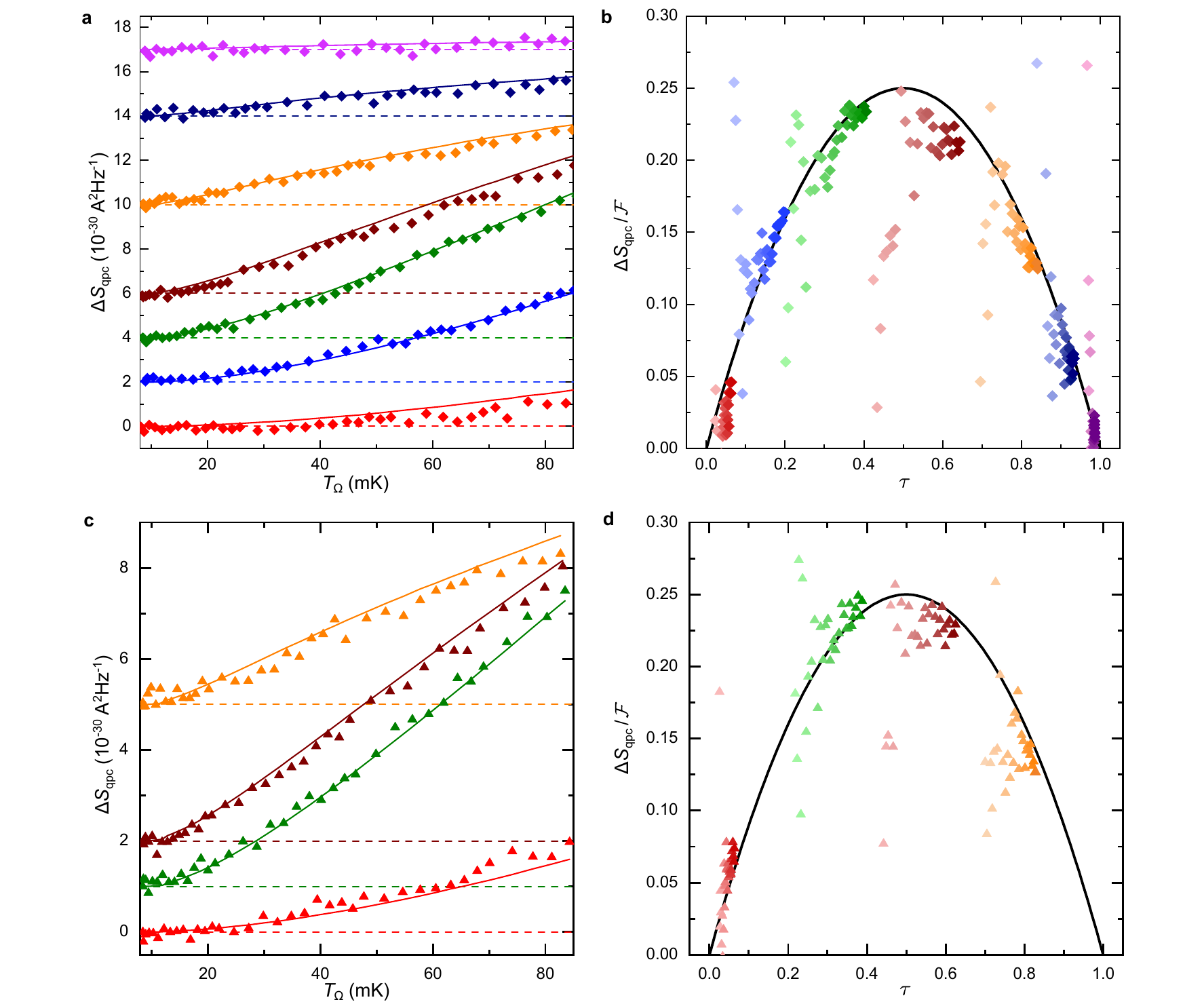}
\footnotesize
\caption{
Thermal shot noise in additional circuit configurations at base temperature $T\simeq8\,$mK.
The thermal shot noise data shown in Fig.~2 represent measurements performed with the circuit tuned at $N=2$.
Here, we display the thermal shot noise and $\Delta S_\mathrm{qpc}/\mathcal{F}$ measured with $N=3$ in panels (a), (b) and $N=4$ in panels (c), (d).
}
\label{figED_SN8mK-Rks34}
\end{figure*}

\begin{figure*}[h]
\renewcommand{\figurename}{\textbf{Supplementary Figure}}
\renewcommand{\thefigure}{\textbf{\arabic{figure}}}
\includegraphics[width=\textwidth]{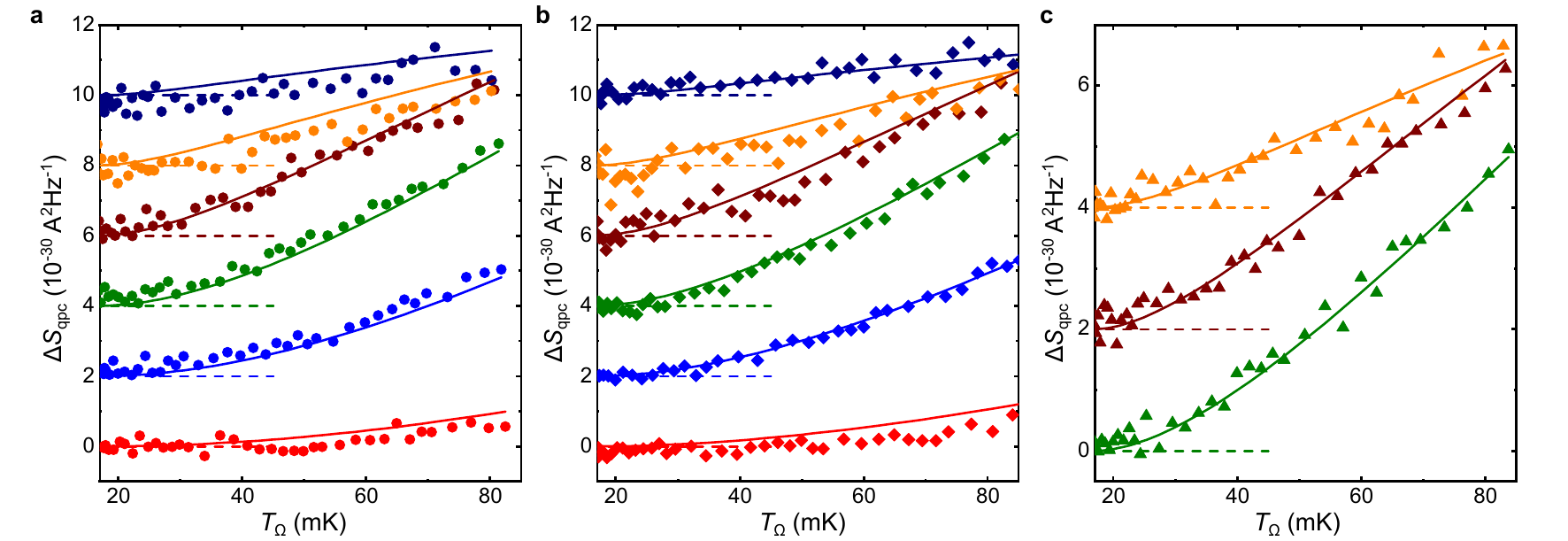}
\footnotesize
\caption{
Thermal shot noise at $T\simeq16\,$mK.
The figure displays similar thermal shot noise data as those shown in Fig.~2a and in Supplementary Fig.~1a,c, but here measured at a larger base temperature $T\simeq16$\,mK.
The panels (a), (b) and (c) display as symbols the experimental thermal shot noise obtained, respectively, in the circuit configurations $N=2,$ 3 and 4.
Continuous lines are the quantitative predictions of Eq.~1, without adjustable parameter.
Different colors represent different QPC settings (same setting as corresponding $T\simeq8\,$mK data, shown with the same color).
Plots are offset for clarity with the offsets displayed as horizontal dashed lines.
}
\label{figED_16mK-Rks234-SN}
\end{figure*}

\begin{figure*}[h]
\renewcommand{\figurename}{\textbf{Supplementary Figure}}
\renewcommand{\thefigure}{\textbf{\arabic{figure}}}
\includegraphics[width=\textwidth]{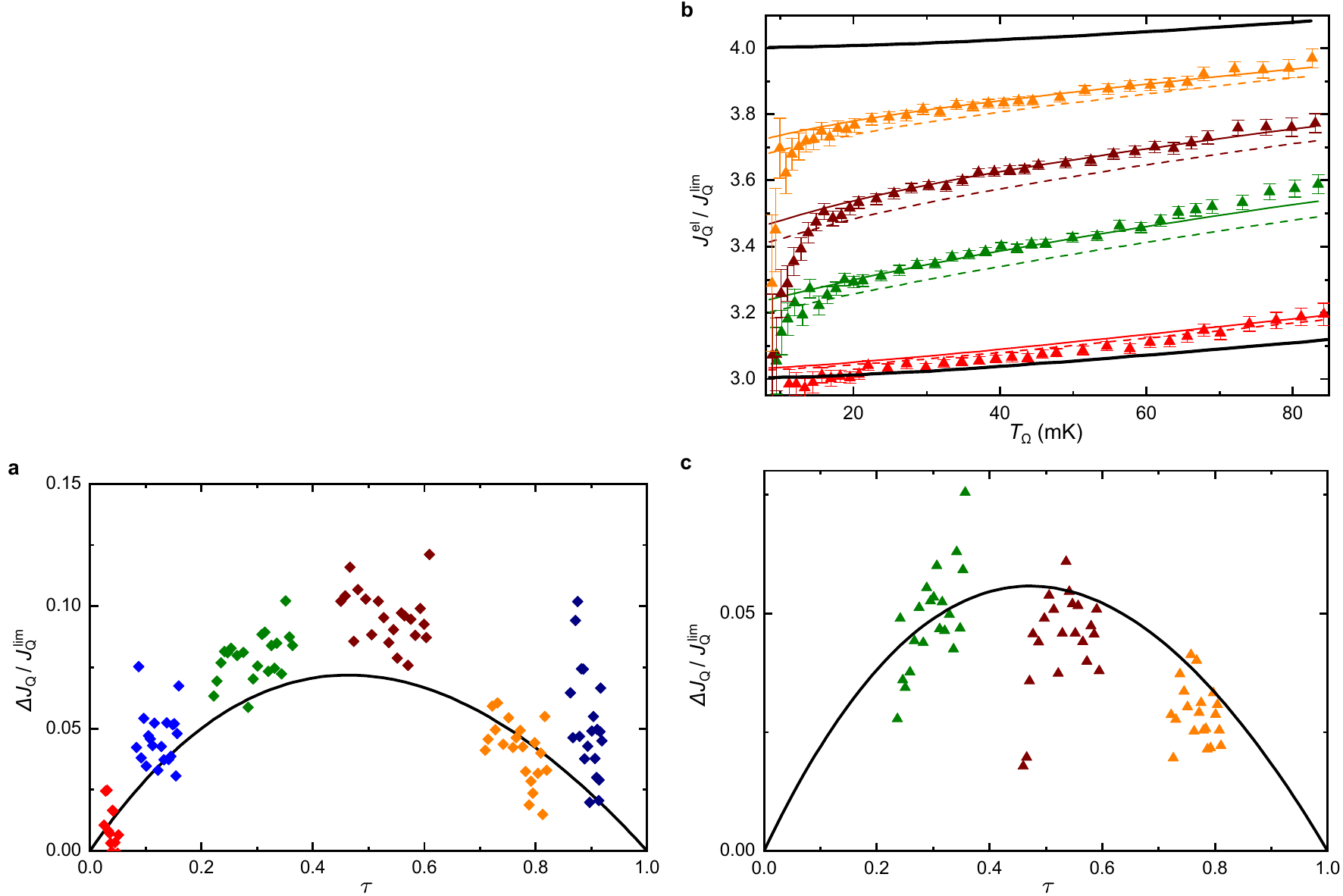}
\footnotesize
\caption{
Electronic heat flow in additional circuit configurations at $T\simeq8\,$mK.
This Supplementary Figure complements Fig.~3 by showing $\Delta J_\mathrm{Q}^\mathrm{el}/J_\mathrm{Q}^\mathrm{lim}$ for $N=3$ (a) and $N=4$ (c), and by showing $J_\mathrm{Q}^\mathrm{el}/J_\mathrm{Q}^\mathrm{lim}$ for $N=4$ (b).
Error bars in (b) represent the standard statistical error.
}
\label{figED_heatFlow-Rks34}
\end{figure*}

\begin{figure*}[h]
\renewcommand{\figurename}{\textbf{Supplementary Figure}}
\renewcommand{\thefigure}{\textbf{\arabic{figure}}}
\includegraphics[width=0.6\textwidth]{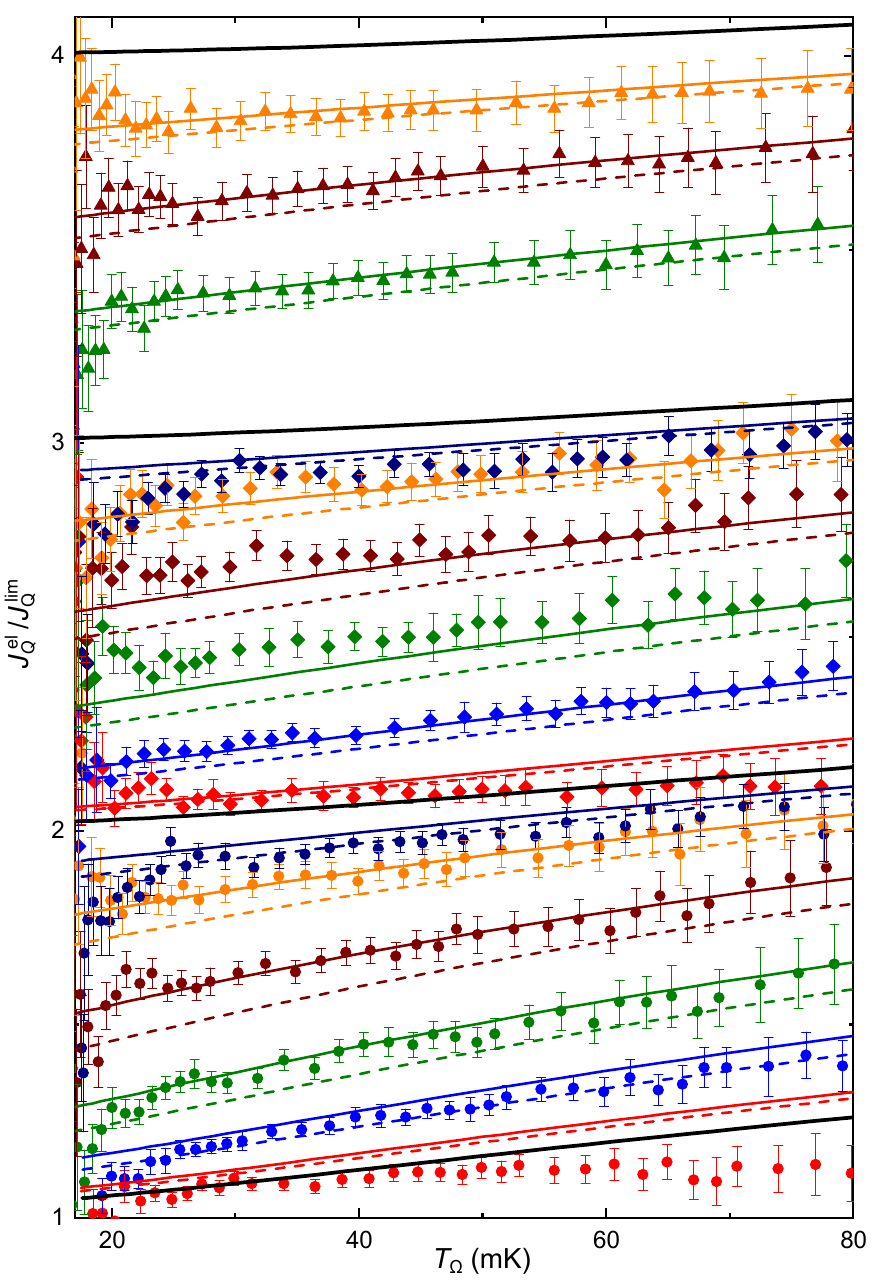}
\footnotesize
\caption{
Electronic heat flow at $T\simeq16\,$mK for different circuit settings ($N\in\{2,3,4\}$) as a function of the temperature $T_\Omega$ of the electrons in the island (similar to the data in Fig.~3a and Supplementary Fig.~3b but at $T\simeq16\,$mK).
The symbols represent experimental measurements with error bars corresponding to the standard statistical error.
The continuous lines show the theoretical predictions of Eq.~23.
The dashed lines are linear interpolations between the two nearest ballistic predictions (black continuous lines) weighted by the simultaneously measured $\tau(T_\Omega)$.
}
\label{figED_16mK-Rks234-JQ}
\end{figure*}

\end{document}